\documentclass{PoS}

\usepackage{amssymb,amsmath,textcomp}
\usepackage{graphicx} 
\usepackage[caption=false]{subfig}	
\usepackage{slashed}

\title{\vspace{-2cm}
{\small \normalfont \hfill DESY 13-193\\
}
\vspace{1cm} Stabilizing the electroweak vacuum by higher dimensional operators in a
Higgs-Yukawa model}

\ShortTitle{Stabilizing the electroweak vacuum by higher dimensional operators in a
Higgs-Yukawa model}

\author{Prasad Hegde\\
       National Taiwan University, Taipei, Taiwan\\
       E-mail: \email{phegde@physik.uni-bielefeld.de}}

\author{Karl Jansen\\
       NIC, Desy Zeuthen, Germany\\
       E-mail: \email{karl.jansen@desy.de}}

\author{C.-J. David Lin \\
       National Chiao Tung University, Hsinchu, Taiwan\\
       E-mail: \email{dlin@mail.nctu.edu.tw}}

\author{\speaker{Attila Nagy}%
        \\
        Humboldt University Berlin; NIC, Desy Zeuthen, Germany\\
        E-mail: \email{nagy@physik.hu-berlin.de}}

\abstract{The Higgs boson discovery at the LHC with a mass of approximately 126~GeV suggests,
that the electroweak vacuum of the standard model may be metastable at 
very high energies.
However, any new physics beyond the standard model can change this picture.
We want to address this important question within a lattice
Higgs-Yukawa model as the limit of the standard model (SM).
In this framework we will probe the effect of a
higher dimensional operator for which we take a $(\phi^{\dagger}\phi)^3$-term.
Such a term could easily originate as a remnant of physics beyond the SM
at very large scales.

As a first step we investigate the phase diagram of the model including
such a $(\phi^{\dagger}\phi)^3$ operator. Exploratory results suggest the existence of
regions in parameter space where first order transitions
turn to second order ones, indicating the existence
of a tri-critical line. 
We will explore the phase structure and the consequences for the stability of the SM, both analytically by investigating the constraint effective potential in
lattice perturbation theory, and by studying the system non-perturbatively using lattice simulations.
}

\FullConference{31st International Symposium on Lattice Field Theory LATTICE 2013\\
                 July 29 -- August 3, 2013\\
                 Mainz, Germany}

\begin{document}
\section{Introduction}
In 2012 a boson was discovered at the LHC with a mass around 126~GeV. 
Experimental results suggest, that this particle is compatible 
with \textit{the} standard model Higgs boson \cite{Aad:2013wqa,Chatrchyan:2013lba}. 
Subsequent analyses following this discovery consolidated this 
interpretation, although a final proof is still missing. 
With the discovery of 
the Higgs boson and the measurement of its mass, 
the standard model (SM) is complete 
in the sense that in 
principle all its parameters are known. 
On the other hand, the SM cannot account for a number of phenomena 
observed in nature such as the quark mass hierarchy, the baryon 
asymmetry of the universe, the amount of CP violation and the 
existence of dark matter. Thus, at some yet unknown energy, the 
SM is expected to be replaced by some more comprehensive theory 
which can, hopefully, explain these physical phenomena.  

The SM itself can provide information on its validity. 
This originates from the fact that the possible mass range of 
the Higgs boson is bounded and that those bounds depend on the cutoff of the theory. 
The upper bound is related to the triviality of the theory, i.e. the 
observation that the quartic coupling runs to zero when the cut-off 
is sent to infinity \cite{Dashen:1983ts}. The lower Higgs boson mass bound is determined by the 
requirement, that the
electroweak vacuum is stable.

Theoretical developments indicate, that a Higgs boson with 
a mass below $\approx$~129~GeV results in a metastable 
vacuum \cite{Degrassi:2012ry}, although this result is still affected
by uncertainties coming mainly from present errors of the strong coupling constant 
and, in particular, the top quark mass. 
The scale at which this metastability occurs can be estimated 
from the evolution of all standard model couplings 
from the electroweak scale up to the Planck scale. 
The metastability then results from the fact, that 
the quartic coupling turns negative at a certain energy scale. 
These calculations are performed solely within the framework 
of the SM and no extensions are considered. 


As a consequence of triviality the cutoff cannot be removed
and the SM needs to be considered as an effective theory only. 
This allows the inclusion of higher dimensional operators
in the theory which can be interpreted as being induced 
by some physics beyond the SM. 
A possible minimal extension of the standard model would be, to add a 
dimension-6 operator, namely a $(\phi^{\dagger} \phi)^3$-term, 
to the scalar part of the standard model. This term   
with a positive coupling $\lambda_6$ 
stabilizes the effective potential even in case of negative 
quartic self coupling.

In this work we investigate the influence of  
such a dimension-6 operator in the framework of a Higgs-Yukawa 
model, which is a reduction 
of the standard model to only consider the complex scalar doublet and quarks. 
For our computations we will use a lattice regularization 
of the Higgs-Yukawa model employing 
a chirally invariant lattice formulation. 
Eventually we want to test, how strongly one can alter the 
lower Higgs boson mass bound and the vacuum structure of the theory, 
if a $(\phi^{\dagger} \phi)^3$-term is present.
Results from an analysis of the renormalization group approach in a $Z_2$-symmetric Higgs-Yukawa model 
suggest, that the mass bound indeed can be decreased \cite{Gies:2013fua}.
%

As a first step we will  
map out the phase structure of the model in the presence of such a 
$\lambda_6\,(\phi^{\dagger} \phi)^3$-term, since the more complicated 
Higgs potential may lead to additional phase transitions which 
can put bounds on the parameters of the theory. 
The exploration of the phase diagram 
will be carried out non-perturbatively 
by means of numerical simulations and the results will be compared to 
(lattice) perturbative calculations
of the constraint effective potential. 

%
\section{Higgs-Yukawa model and implementation}
The field content of the Higgs-Yukawa model is given by a  
fermion doublet $\Psi=(t, b)^T$ and the scalar complex doublet $\varphi$. 
We restrict ourselves to the simple case of one fermion 
doublet with mass degenerate quarks even though in principle a more general case
would be possible but significantly more computertime expensive. 
In the continuum notation the action is given by:
\begin{multline}\label{eq:action_continuum}
 S^{\text{cont}}[\bar{\psi}, \psi, \varphi] = \int d^4 x \left\{\frac{1}{2}\left(\partial_{\mu} \varphi \right)^{\dagger} \left(\partial^{\mu} \varphi \right)
			+  \frac{1}{2} m_0^2 \varphi^{\dagger} \varphi 
			+ \lambda \left(\varphi^{\dagger} \varphi \right)^2 
			+ \lambda_6 \left(\varphi^{\dagger} \varphi \right)^3 \right\} \\
			+\int d^4 x  \left\{\bar{t} \slashed \partial t + \bar{b} \slashed \partial b +
			y \left( \bar{\psi}_L \varphi\, {b}_{_R} + \bar{\psi}_L \tilde \varphi\, {t}_{_R} \right)
			+ h.c. \right\},
\end{multline}
with $\tilde \varphi = i\tau_2\phi^{*}$ and $\tau_2$ being the second Pauli matrix. The bare standard model parameters 
are given by $m_0^2$ and $\lambda$ for the Higgs potential and $y$ for the Yukawa coupling. 
Further, we added the dimension-6 operator $\lambda_6 \left(\varphi^{\dagger} \varphi \right)^3$ 
to the action. 

For the numerical implementation of this model we us a polynomial 
hybrid Monte Carlo algorithm~\cite{Frezzotti:1998eu} with dynamical overlap 
fermions. Details of the implementation can be found in~\cite{Gerhold:2010wy}. 
On the lattice, it is convenient to rewrite the bosonic part of the action 
in an Ising model like way:
\begin{equation}\label{eq:bosonic_action_lattice}
S_B[\phi] = -\kappa \sum\limits_{x,\mu} \phi_x^{\dagger} \left[\phi_{x+\mu} + \phi_{x-\mu}\right] + 
					\sum\limits_{x} \left( 
					\phi_x^{\dagger} \phi_x + 
					\hat{\lambda} \left[ \phi_x^{\dagger} \phi_x - 1 \right]^2 + 
					\hat{\lambda}_6 \left[ \phi_x^{\dagger} \phi_x \right]^3 \right).
\end{equation}
Here the scalar field is represented as a real four-vector and the 
relation to the continuum notation is given by:
\begin{equation}\label{eq:rescaling_cont_lattice}
 \varphi(x) = \sqrt{2 \kappa} \left( \begin{array}{c} \phi_x^2 + i\phi_x^1 \\ \phi_x^0 - i \phi_x^3 \end{array} \right) ,\quad
	 m_0^2 = \frac{1 - 2 \hat{\lambda} -8 \kappa}{\kappa} ,\quad
	 \lambda = \frac{\hat{\lambda}}{{4 \kappa^2}},\quad
	 \lambda_6 = \frac{\hat{\lambda}_6}{{8 \kappa^3}}.
\end{equation}
For the determination of the phase structure, we employ the 
magnetization $m$ as order parameter. The magnetization is given by the
modulus of the average scalar field and is related to its 
vacuum expectation value ($vev$) via:
\begin{equation}\label{eq:mag_and_vev}
 m = \left< \left | \frac{1}{V}\sum\limits_x \phi_x \right| \right>,\qquad
 vev = \sqrt{2 \kappa} \cdot m.
\end{equation}

\section{Constraint effective potential}
To compare our numerical results to perturbation theory we 
employ the constraint effective potential (CEP) \cite{Fukuda:1974ey, O'Raifeartaigh:1986hi}. 
The basic idea is that the potential $U(\hat v)$ that only depends on 
the zero mode $\hat \nu$ of the scalar field corresponding to its
vacuum expectation value ($vev$).
If one assumes that the groundstate of the system is at a non-vanishing $vev$, 
the scalar doublet can be decomposed into the Higgs mode and 
three goldstone modes. 
The perturbative calculations are done by explicitly keeping the lattice 
regularisation, i.e.\ for the fermionic determinant the overlap operator 
is used and all sums over lattice momenta are performed numerically. 
A derivation of the constrained effective potential used here 
can be found in \cite{Gerhold:2007gx}.
The CEP up to the first order in $\lambda$ and $\lambda_6$ is given by:
\begin{multline}\label{eq:CEP_with_phi_6}
 U(\hat v) = U_f(\hat v) + \frac{m_0^2}{2} {\hat v}^2 +\lambda {\hat v}^4 + \lambda_6 {\hat v}^6 \\
                         + \lambda \cdot {\hat v}^2 \cdot 6(P_H+P_G)
                         + \lambda_6 \cdot \left( {\hat v}^2 \cdot ( 45 P_H^2 + 54 P_G P_H + 45 P_G^2)
                         + {\hat v}^4 \cdot ( 15 P_H + 9 P_G ) \right ).
\end{multline}
The fermionic contribution $U_f$ comes from integrating out the 
fermions in the background of a constant field:
\begin{equation} \label{eq:fermionic_contribution_CEP_massbound}
 U_f(\hat v) = -\frac{4}{V} \sum\limits_p \log\left| \nu(p) + y \cdot \hat v \cdot \left( 1-\frac{1}{2 \rho} \right) \nu(p) \right|^2,
\end{equation}
with $\nu(p)$ denoting the eigenvalues of the overlap operator corresponding to the momentum $p$:
\begin{equation} \label{eq:eigenvalues_of_overlap}
 \nu(p) = \rho \left( 1 + \frac{ i \sqrt{{\tilde p} ^2} + r  {\hat p}^2 - \rho}
                           {\sqrt{ {\tilde p} ^2 + \left( r  {\hat p}^2 - \rho\right)^2}}\right),\qquad
 {\hat p}^2 = 4 \sum\limits_{\mu} \sin^2\left(\frac{p_{\mu}}{2}\right),\qquad
 {\tilde p}^2 = \sum\limits_{\mu} \sin^2\left(p_{\mu}\right).
\end{equation}
The propagator sums for the Higgs and the Goldstone bosons are given by:
\begin{equation} \label{eq:def_propagator_sums}
 P_{H} = \frac{1}{V} \sum\limits_{p \neq 0} \frac{1}{{\hat p}^2 + m_{H}^2} ,\qquad
 P_{G} = \frac{1}{V} \sum\limits_{p \neq 0} \frac{1}{{\hat p}^2}.
\end{equation}
We note, that even though we use the continuum notation, 
all quantities are meant to be dimensionless, i.e. the lattice spacing
is set to one implicitly.
The $vev$ is given by the minimum of the potential, 
and by setting it to the phenomenologically known value of $vev$ of 246~GeV 
one obtains a physical scale in this approach:
\begin{equation}\label{eq:vev_and_cutoff_CEP}
 \left. \frac{\text{d}U(\hat v)}{\text{d}\hat v} \right|_{\hat v = vev} \stackrel{!}{=} 0 ,\qquad
 \Lambda = \frac{246 \text{ GeV}}{vev}.
\end{equation}
Further, the squared Higgs boson mass $m_H^2$ is determined by the 
curvature of the potential at its minimum:
\begin{equation}\label{eq:mhSquared_from_CEP}
 \left. \frac{\text{d}^2 U(\hat v)}{\text{d}{\hat v}^2} \right|_{\hat v = vev} = m_H^2.
\end{equation}
Due to the explicit appearence of the Higgs boson mass in the 
propagator sum \eqref{eq:def_propagator_sums}, we have to use an 
iterative approach to solve the CEP. To this end, we fix the parameters 
$m_0^2$, $y$, $\lambda$ and $\lambda_6$
and iterate  eqs.~(\ref{eq:vev_and_cutoff_CEP},\ref{eq:mhSquared_from_CEP}) until
we find convergence. 


%
\section{Results}
As already mentioned, this work is a first step towards a 
systematic investigation whether it is possible to alter 
the lower Higgs boson mass
bound if in addition to the ordinary standard model Higgs potential a 
dimension-6 operator $\lambda_6 (\phi^{\dagger} \phi)^3$ is included. To this end, 
we have to map out
the bulk (non-thermal) phase structure.
The question is where phase 
transitions occur and of which order these phase transitons are.
In order to separate the cut-off scale from the low-energy quantities, 
we have to search for 2nd order phase transitions.  On the other hand, 
it is natural to expect the appearance of 1st order phase transitions 
in the presence of the dimension-6 operator.  We would like to be 
certain that our study of the Higgs-boson mass bounds is performed away 
from these 1st order phase transitions.
Locating the tri-critical line in the $\lambda$-$\lambda_6$ plane is then 
the first task of our investigation.  
In all calculations we will 
keep the Yukawa couplings fixed to obtain the physical top quark mass 
for which we follow ref.~\cite{Gerhold:2009ub} and use 
the tree-level relation
\begin{equation}\label{eq:treeLevel_topMass}
 y = \frac{m_t}{vev} = \frac{175 \text{ GeV}}{246 \text{ GeV}}.
\end{equation}

For our study we have employed two values of the coupling 
$\lambda_6$ and for each we choose a set of values for 
$\lambda \leq 0$. With the Yukawa coupling
fixed according to 
\eqref{eq:treeLevel_topMass}\footnote{
The couplings $\lambda$ and $\lambda_6$ in this section are 
the dimension-less couplings in the 
continuum notation and may not to be confused with the parameters 
being rescaled by powers of $\sqrt{2\, \kappa}$},
there is only one free parameter left, namely $m_0^2$ which is 
directly related to $\kappa$ according to \eqref{eq:rescaling_cont_lattice}. 
We then perform scans in $\kappa$  
to determine regions of broken and symmetric phases corresponding to 
clearly non-zero and almost zero values of the magnetization.

\begin{figure}[htb]\centering
	 \centering
	 \subfloat[$\lambda_6=0.10$]{\includegraphics[width=0.45\textwidth]{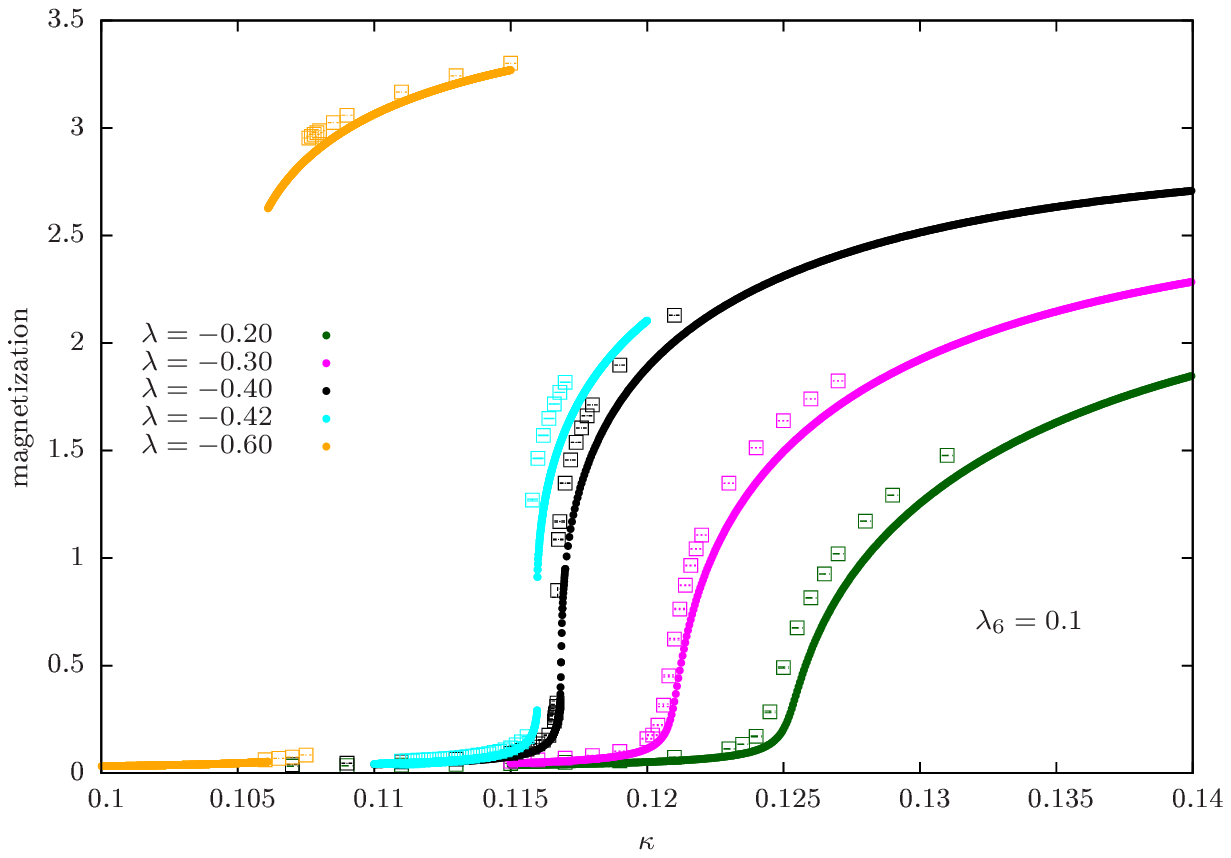}\label{fig:CEP_vs_sim_0p1}}
	 \subfloat[$\lambda_6=1.00$]{\includegraphics[width=0.45\textwidth]{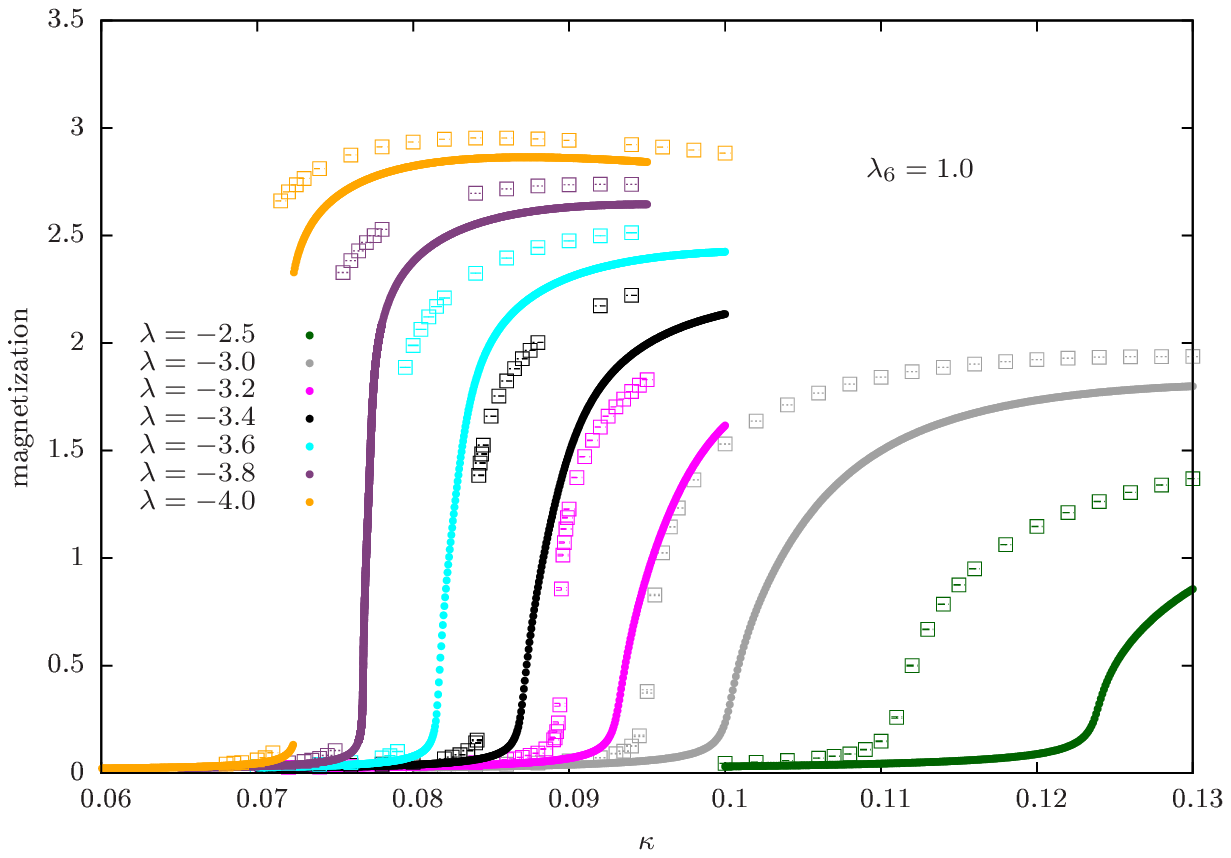}}
	 \caption{
These plots show the magnetization as it is obtained in 
simulations (open squares) and from the CEP (dots) for 
two values of $\lambda_6$. 
Both, the simulations and the evaluation of the CEP were 
done on $12^3\times 24$ lattices. The left plot shows results for a rather small value of 
$\lambda_6=0.1$ and the agreements between perturbative and 
non-perturbative determination is quite good, whereas the 
agreement for the non-perturbative
case of $\lambda_6=1$ can only be considered as qualitative. 
	 }
	 \label{fig:CEP_vs_sim}
\end{figure}

In figure~\ref{fig:CEP_vs_sim} we show results comparing curves 
obtained from numerical evaluations of the CEP \eqref{eq:CEP_with_phi_6} 
and direct numerical simulations
performed on rather small lattices of volume $V=12^3\times 24$. 
For both tested values of $\lambda_6$ depending on the choice of 
$\lambda$ there is quite clear evidence 
for either second or first order phase transitions.

\begin{figure}
\centering
\subfloat[trajectories]{\includegraphics[width=0.45\linewidth]
{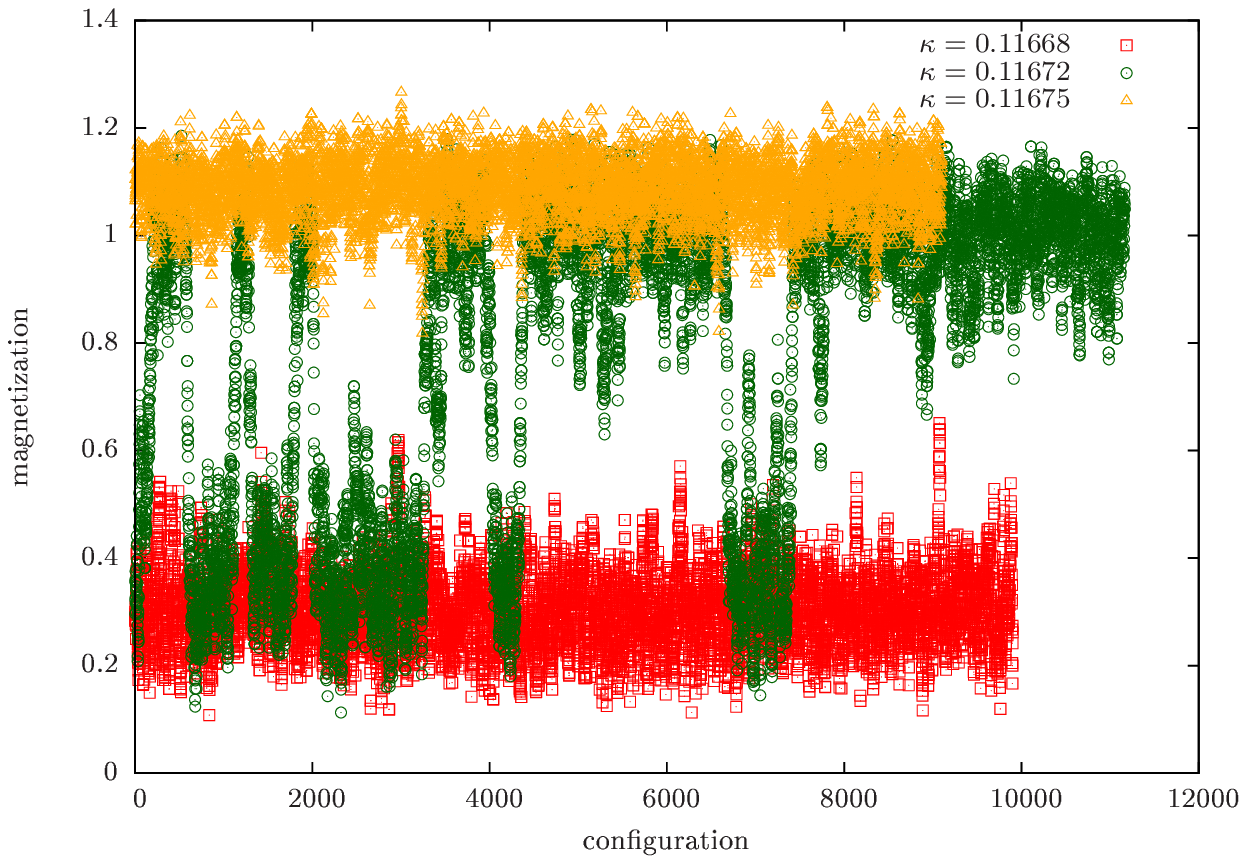}}
\subfloat[CEP from simulation]{\includegraphics[width=0.45\linewidth]
{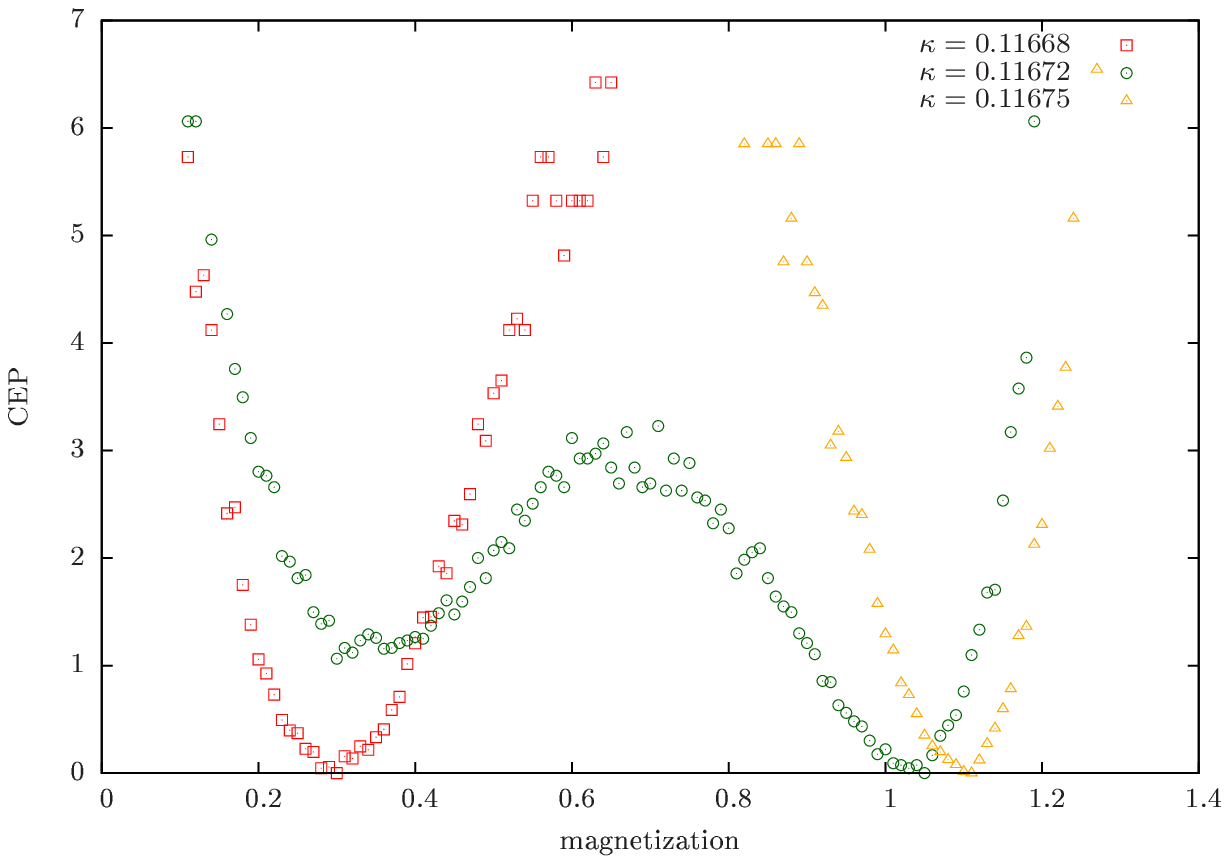}}
\caption{The left plot shows the Monte Carlo time trajectories of the magnetization 
corresponding to simulation data for 
$\lambda=-0.4$ from fig.~\protect\ref{fig:CEP_vs_sim_0p1}
where one observes a typical metastable behaviour for $\kappa=0.11672$ with 
the magnetization jumping between two values. The runs of adjacent $\kappa$ do not 
show this behaviour.
The right plot shows the CEP as it is obtained from the simulation
for those three values of $\kappa$. Both plots nicely indicate 
the existence of a first order phase transition. 
}
\label{fig:PT_from_simulation}
\end{figure}

To test, whether some of the transitions we observe are  
of first order, we show the Monte Carlo time trajectories for the 
magnetization for a chosen set of 
$\kappa$-values in figure~\ref{fig:PT_from_simulation}. There 
three neighboring kappa values are shown and one can observe, 
that while the smallest and the largest values of $\kappa$ lead 
to stable runs,
the middle value shows metastabilities where the system 
jumps between two values of the magnetization. 
It is also possible, to access the CEP from the 
simulation data. Up to a constant, the CEP can be related to the 
logarithm of the histogram of the  
magnetization, see e.g.~\cite{Dimitrovic:1991qg}.

\begin{figure}
\centering
\subfloat{\includegraphics[width=0.45\linewidth]
{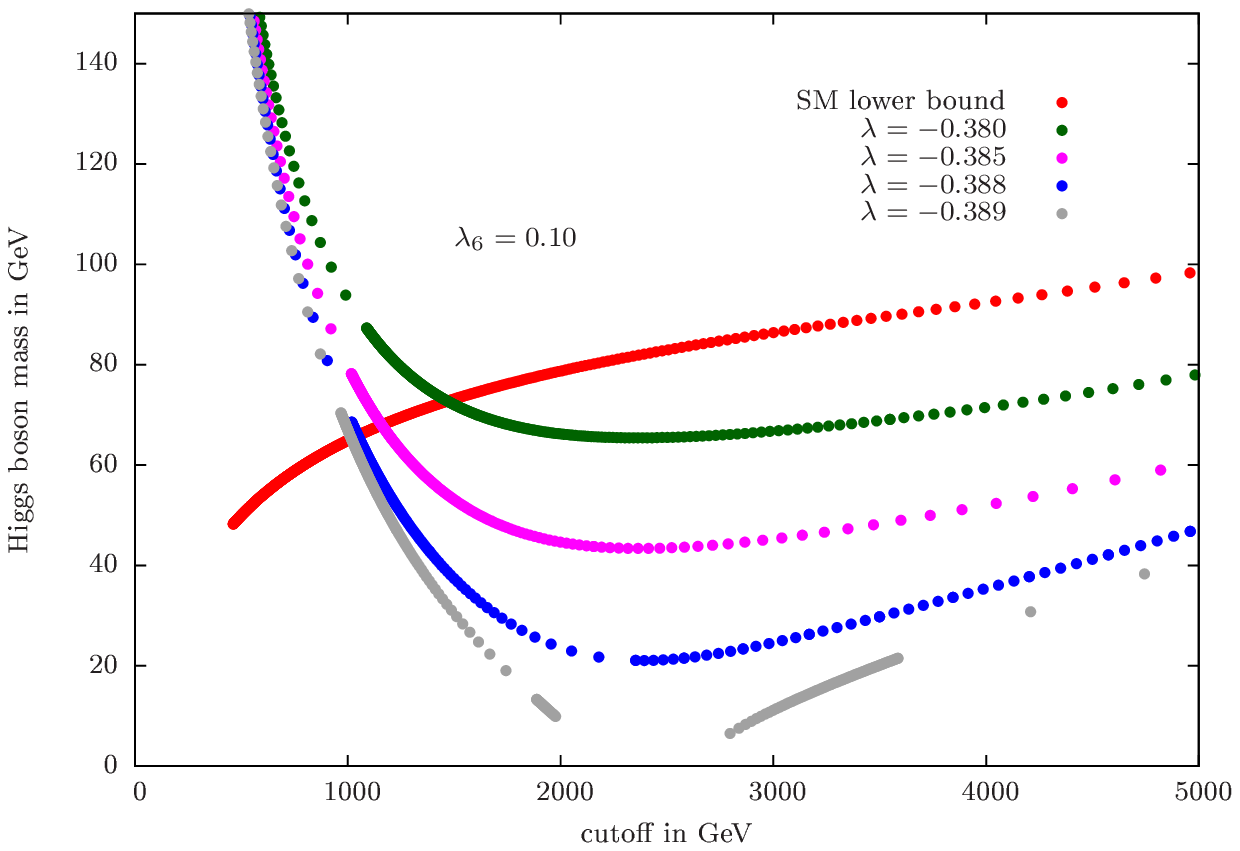}}
\subfloat{\includegraphics[width=0.45\linewidth]
{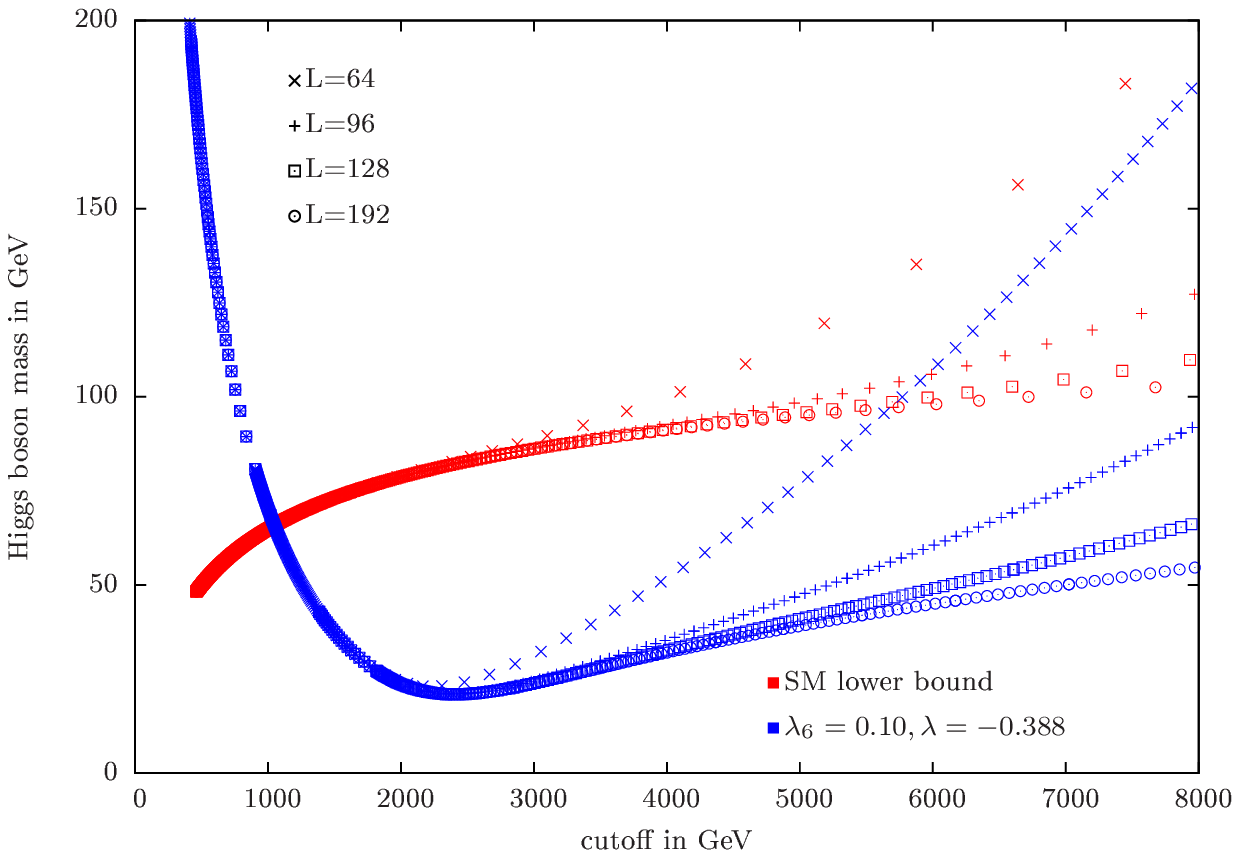}}
\caption{
Here we show the dependence of the Higgs boson mass 
on the cutoff as it is obtained in the CEP. The left plot compares 
the results for various values of $\lambda$ on a $96^3\times192$ 
lattice while keeping 
$\lambda_6=0.1$ constant. Additionally  we show the 
standard model lower bound indicated by the red points.
The gap in the gray data points  originates from the first order 
phase transition. 
The right plot shows the volume dependence for various $L^3 \times 2\,L$ 
lattices while keeping $\lambda=-0.388$ and $\lambda_6=0.1$ fixed. 
It also shows the volume dependence 
of the standard model mass bound.}
\label{fig:mass_vs_cutoff}
\end{figure}

To address the question how the addition of the $\lambda_6$-term 
alters the Higgs mass bounds, we are so far restricted to the 
mass determination from the CEP.
The standard model
lower Higgs boson mass bound is obtained by setting $\lambda$ to zero~\cite{Gerhold:2009ub}.
If a $\lambda_6$ coupling is switched on, we find a behaviour 
of the Higgs boson mass as 
shown in figure~\ref{fig:mass_vs_cutoff} for $\lambda_6=0.05\text{ and } 0.1$. 
In both cases, it is possible, to find (negative) values for $\lambda$, that allow 
a significant decrease of the Higgs boson mass for intermediate 
cutoffs without entering the region of first order phase transitions 
discussed above. 
In fig.~\ref{fig:mass_vs_cutoff} we show finite volume, but also 
the infinite volume curves demonstrating that 
that the shift in 
the mass survives the infinite volume limit.

\section{Conclusions and outlook}
In this work, we have added a 
dimension-6 operator 
to a Higgs-Yukawa model to test the stability of a so extended
SM.  
We found that for fixed values of $\lambda_6 = 0.1$  and for a cutoff 
of about~$\gtrsim 1.5 \text{TeV}$, the Higgs boson mass can be lowered when the quartic coupling 
is driven more and more negative, as was also 
found in ref.~\cite{Gies:2013fua}. In addition, we detected that for a certain 
(negative) value of the quartic coupling the transition between the 
symmetric and the broken phase turns first order 
and the separation between the cut-off and the low-energy scale is lacking, leading to
an absolute lower bound 
of the Higgs boson mass. 
With this we conclude that for the here considered value
of a $\lambda_6$ coupling  
a Higgs boson mass of 126GeV is fully compatibale with an addition 
of a $\phi^6$ term. In addition, since such a term is quadratically 
suppressed with the cutoff, it has no effect at very high energies.

An interesting question remains, however, namely what happens at 
very large values of $\lambda_6=O(10)$. Will there be clash 
with the Higgs boson mass value leading to bounds of the 
$\lambda_6$ coupling like in the case of our fourth fermion 
investigation \cite{Bulava:2013ep}. Or, can it even be possible that are a 
low values of the cutoff of O(10TeV) a metastable behaviour can be found? 
We plan to address these question in the future by studying 
larger values of the $\lambda_6$ coupling.

\section{Acknowledgement}
Simulations have been performed at the SGI system HLRN-II at the HLRN supercomputing service Berlin-Hannover and
the PAX cluster at DESY-Zeuthen. This work is supported by
Taiwanese NSC via the grant 102-2113-M-009-002-MY3  and by the DFG through the DFG-project Mu932/4-4.
%
%

\end{document}